# Term-Class-Max-Support (TCMS): A Simple Text Document Categorization Approach Using Term-Class Relevance Measure

D S Guru
Department of Studies in Computer Science
University of Mysore, Manasagangotri
Mysore, INDIA
dsg@compsci.uni-mysore.ac.in

Mahamad Suhil
Department of Studies in Computer Science
University of Mysore, Manasagangotri
Mysore, INDIA
mahamad45@yahoo.co.in

*Abstract*—In this paper, a simple text categorization method using term-class relevance is proposed. Initially, text documents are processed to extract significant terms present in them. For every term extracted from a document, we compute its importance in preserving the content of a class through a novel term-weighting scheme known as Term-Class Relevance (*TCR*) measure proposed by Guru and Suhil (2015). In this way, for every term, its relevance for all the classes present in the corpus is computed and stored in the knowledgebase. During testing, the terms present in the test document are extracted and the term-class relevance of each term is obtained from the stored knowledgebase. To achieve quick search of term weights, B-tree indexing data structure has been adapted. Finally, the class which receives maximum support in terms of term-class relevance is decided to be the class of the given test document. The proposed method works in logarithmic complexity in testing time and simple to implement when compared to any other text categorization techniques available in literature. The experiments conducted on various benchmarking datasets have revealed that the performance of the proposed method is satisfactory and encouraging.

## I. INTRODUCTION

Due to the drastic increase in the amount of text content over the internet, it has become a crucial task to design efficient systems to process and manipulate such data to infer useful results (Sebastiani, 2002, Lam et al., 1999). Text categorization (TC) is the process of automatically classifying a given text document into one of the many predefined categories. It carries higher importance due to its huge impact on subsequent activities of text mining and also due to many applications involving text categorization such as spam filtering in emails, classification of medical documents, sentiment analysis etc., (Harish et al., 2010, Aggarwal and Zhai, 2012).

From the literature we can understand that, the effort to design systems for automatic text categorization has the history of more than two decades (Hotho et al., 2005; Aggarwal and Zhai, 2012). With the development of machine learning approaches, there are plenty of techniques proposed for various tasks of TC such as representation, feature selection and categorization. However, the methods either involve higher complexities or they will be less accurate.

Machine learning based TC systems carry the following general structure. All the training documents are preprocessed using stemming, pruning, stopwords removal to retain only content terms. Then a matrix representation to the entire training data is given using vector space model which uses the bag of words (terms) (Li and Jain, 1998; Rigutini, 2004). The dimension of such a matrix will be very high even for a dataset of reasonable size which making the learning algorithms less effective. Hence, dimensionality reduction has been widely explored on text data as a mandatory step in the design of TC to which not only reduce the dimension but also to increase the classification performance (Guyon and Elisseeff, 2003). Most of the works reported in literature of TC have used either feature selection through ranking or feature extraction through transformation as the means of dimensionality reduction. In view of this, we can find a big list of feature selection techniques developed for TC which also include those inherited from Information Retrieval domain for feature ranking such as information gain, gain ratio, Chi-squared statistic, document frequency, *tf*idf*, mutual information, distinguishing feature selector (DFS), gini index, odds ratio etc.,(Yang and Pedersen, 1997; Forman, 2003; Montanes et al., 2005; Javed et al., 2015) and also many works have proposed transformation based dimensionality reduction including PCA, ICA, LSI, non-linear embedding etc., (Makrehchi., 2007; Deerwester et al., 1990; Sebastiani., 2002; Cai et al., 2005; Cai and He 2012; Uysal and Gunal., 2014). Some methods integrate both feature selection and feature extraction to achieve redundancy elimination (Bharti and Singh., 2015). Feature selection techniques have been studied both locally and globally respectively for selecting features for each class then aggregating to form a single set of features and selecting the features for entire data globally independent of class. Finally, a classifier is trained and evaluated with the small set of features obtained after dimensionality reduction (Sebastiani., 2002). Thus, it is a very long and time consuming process. However, we can find many applications where the processing capability is less and moderate classification accuracy is acceptable. For such type of applications, it is essential to design a simple yet effective TC system which can predict the probable class of a test document quickly.

In this paper, we present a simple TC system which predicts the class of an unknown document by estimating the total support of the terms present in the document. To estimate the weight of a term, we search through an indexed knowledge base of term weights which is created during the training stage by the use of B-tree. The term weight that we compute is the weight of a term with respect to a given class using the term_class relevance (*TCR*) measure proposed by Guru and Suhil(2015). The method works with logarithmic complexity irrespective of the size of the corpus.

The rest of the paper is organized as follows: Section 2 describes the proposed TC system along with the *TCR* measure used for term weighting. Experimental setup, datasets, results of the proposed method and analysis are presented in section 3. Finally, section 4 concludes the paper.

## II. THE PROPOSED TC METHOD: TERM-CLASS-MAX-SUPPORT (TCMS)

In this section, we propose a new text classification method in detail. Initially, we present the creation of knowledgebase of term weights using the term_class relevance measure proposed by Guru and Suhil (2015) from the pool of training documents. Then, we propose to index the term weights using B-tree indexing technique for quick retrieval during testing. Finally, when a test document is given, the weights of its respective terms are searched in the indexed knowledgebase and it is classified to the class which attains maximum support by the terms. Fig. 1 depicts the different stages involved in our method.

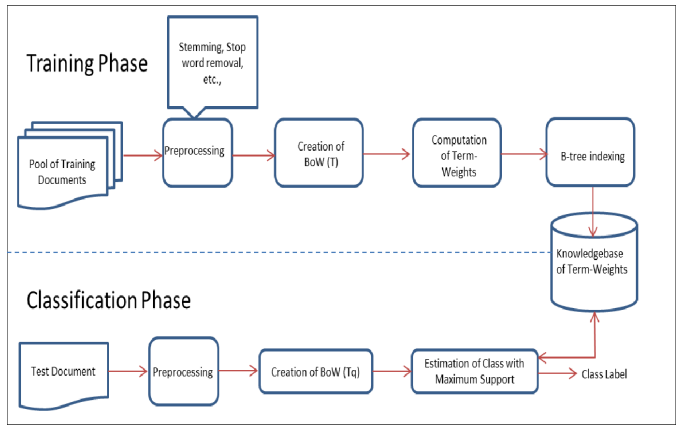

Fig. 1. Graphical representation of the different stages involved in the proposed model

### A. Creation of B-tree indexed knowledgebase of term weights using term_class relevance

Consider a training collection of $N$ labeled documents $D_1, D_2, \ldots, D_N$ with $K$ classes $C_1, C_2, \ldots, C_K$. Initially, preprocessing is applied to all the documents to create a Bag of Words (BoW) of the training collection. During preprocessing, we apply tokenization, stemming and stop word removal convert a document into a set of content terms. Let BoW created for the training set contain d unique terms say $T = \{t_1, t_2, \ldots, t_d\}$. Then we compute the weight of each term $t_i$ in preserving the content of a class using the term_class relevance (TCR) measure proposed by Guru and Suhil (2015). Term_class relevancy is defined as the ability of a term $t_i$ in classifying a document $D$ as a member of a class $C_j$ as given in (1).

$$Term\_Class\,Relevancy(t_i, C_j) = c \times Class\_TermWeight(t_i, C_j) \times Class\_TermDensity(t_i, C_j) \quad (1)$$

Where c is the proportionality constant defined as the weight of the class $C_j$ as given in (2). Class_TermWeight and Class_TermDensity are respectively the weight and density of $t_j$ with respect to the class $C_j$ which are computed using equation (3) and (4) respectively.

$$ClassWeight(C_j) = \frac{\#Documents\,in\,C_j}{\#Documents\,in\,Training\,Set} \quad (2)$$

$$Class\_TermWeight(t_i, C_j) = \frac{\#documents\,in\,C_j\,containing\,t_i}{\#documents\,in\,the\,training\,set\,containing\,t_i} \quad (3)$$

$$Class\_TermDensity(t_i, C_j) = \frac{\#occurences\,of\,t_i\,in\,C_j}{\#occurences\,of\,t_i\,in\,the\,training\,collection} \quad (4)$$

The advantage of using *TCR* against any other term-weighting schemes is that, it directly computes the relevancy of the term with respect to a class of interest so that it can be used as a clue to identify the possible class to which a document may belong without the need of a classifier. The measure uses class as well as corpus information together as opposed to the conventional *tf-idf* scheme, which utilizes the document frequency from only the corpus. This helps in properly deciding the weight of a term without any bias towards a particular class, which in turn helps in deciding the class for a classifier.

The terms obtained during training and their weights computed using *TCR* are stored in a B-tree indexed knowledgebase for quick retrieval during classification stage. We create a B-tree of order say r where every node will have a maximum of *r*-1 elements which are the terms to be indexed and *r* child nodes. Our intension is to have an indexing scheme which helps us in fetching the weight of a term with respect to a class of interest. Hence, along with a term $t_i$, we also preserve its weight to the *K* different classes in the collection as an array $W_i$ and a pointer to it is also stored in the corresponding node of the B-tree. For instance, Fig. 2 shows a node of a B-tree of order 3 which has two terms {term$_1$, term$_2$}, three child node pointers {c$_1$, c$_2$, c$_3$} and two pointers {p$_1$, p$_2$} pointing to weight vectors $W_1$ and $W_2$ respectively.

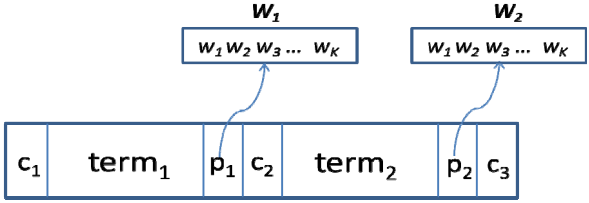

Fig. 2. An example of a B-tree node structure preserving terms and their weights

*B. Classification*

Classification part of our model is very simple. When a test document $D_q$ is given for classification, we follow the same steps as applied for training documents to obtain its content terms say $T_q = \{t_1^q, t_2^q, ..., t_m^q\}$. Then, we estimate the total support ($TS_j$) of $D_q$ for a class $C_j$ by computing the sum of the weights of the terms present in $T \cap T_q$ as given by (5) below.

$$TS_j = \sum_{\forall t_i^q \in T \cap T_q} tf_i^q \times TCR(t_i^q, C_j) \quad (5)$$

where, $tf_i^q$ is the frequency of ith term in the query document Dq which is multiplied by the weight of ith term with respect to $j^{th}$ class $C_j$. In this process, instead of computing the term weights again, we search them from the B-tree indexed knowledgebase. Then, we classify $D_q$ to the class which receives highest support by the terms in $T \cap T_q$ as given in (6) below.

$$result = arg \max_{j=1}^{K} (TS_j) \quad (6)$$

As we search the terms in B-tree, the time required for searching of a term is O(log d) in worst case. Once we search the required term, its weight to K different classes can be accessed through the pointer present along with the term as shown in Fig. 2 which is a linear search with complexity O(K). So, the total time required for classifying a test document with m terms is $Time \propto O(m \times (K + \log d))$.

### III. EXPERIMENTATION

To validate the applicability and effectiveness of the proposed model, we have conducted experiments on three different benchmarking datasets. The performance of the method is evaluated by using the well-known metrics such as Precision, Recall and F-measures. We have computed both macro and micro averaged measures to test the performance on both balanced as well as imbalanced datasets. The three datasets used are 20Newsgroups, Reuters21578 and RCV1-v2.

20Newsgroups (http://qwone.com/~jason/20Newsgroups/) is a data collection of news articles containing18846 documents distributed into 20 different categories in a balanced way. The Reuters21578 dataset contains news articles taken from Reuters with 21578 documents from 135 categories (http://www.daviddlewis.com/resources/testcollections/reuters21578/). It is the most imbalanced dataset available for text categorization. For our experiments we eliminated all such classes whose number of documents are less than 5 which resulted with 8243 documents from 45 classes. Further, we also conducted experiments on RCV1-v2, a benchmarking text dataset from Reuters with 804414 documents. (http://about.reuters.com/researchandstandards/corpus/available.asp). It is a hierarchical dataset with 4 higher level categories which can be subdivided into 103 subcategories. In our experiments we have considered a subset of the whole dataset with 4 higher level categories consisting of 9625 randomly chosen documents. Fig. 3 shows the distribution of documents in different classes of each dataset.

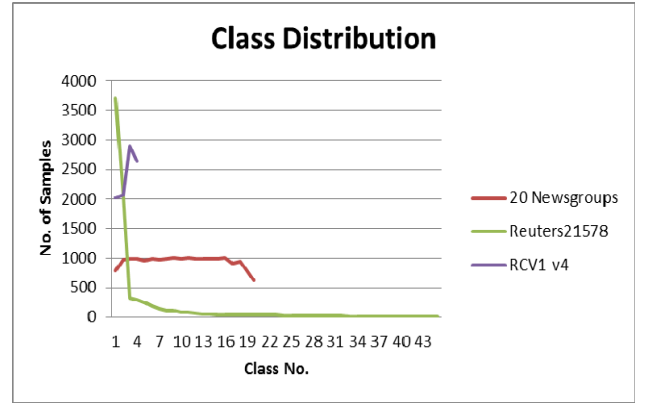

Fig. 3. Distribution of samples in different classes of various datasets

Experiments were conducted on the datasets with different percentages of training samples varying from 10 to 80 percent. The average performance of the proposed model for 10 different random trials has been plotted in Fig. 4. It can be observed from Fig. 4 that, for 20 Newsgroups and RCV1-v2 datasets, the performance has been consistently improved with the increase in the amount of training data. In case of Reuters-21578, the value of micro-F has seen a gradual increase as we increased the training percentage up to 70 and it has suddenly dropped after that point. But, the value of macro-F is too low and has seen no significant rise with the increase in the percentage of training data. The main reason for this is the skewness of Reuters21578 dataset as it can be seen from Fig. 3. Though the method is underperforming when it comes to its competitiveness with the state of art techniques, we suggest it could be used as a tool to come out with initial guesses of the probable classes for a given test document. Hence, given a test document, instead of working with all K classes present in the population, a classifier designed for classification can take only a subset of top

$K'$ classes sorted according to their support generated by the terms present in the test document.

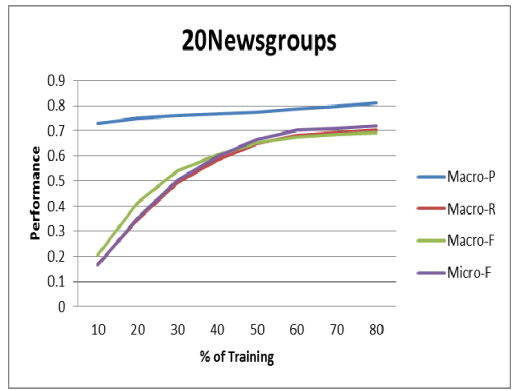

(a) 20-Newsgroups

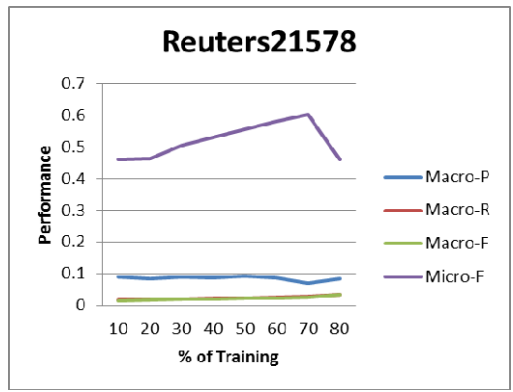

(b) Reuters21578

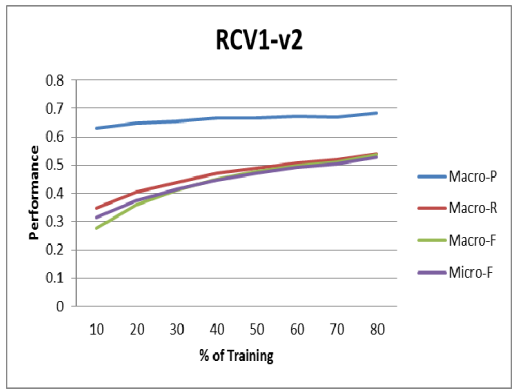

(c) RCV1-v2

Fig. 4. Performance of the proposed method in terms of macro and micro measures for different datsets.

## IV. CONCLUSIONS

In this paper, we have proposed a simple text categorization method using *TCR* measure, a term-weighting scheme which measures the relevance of a term with respect to a given class. A B-tree indexed knowledgebase of term-weights computed using *TCR* measure is created for the terms present in the training corpus. During testing, the term-weights of every term in a test document is fetched from the knowledgebase with respect to the different classes. The sum of the weights of the terms present in the test document with respect to a particular class is considered to be the support of the document to the respective class. Hence, a class which receives maximum support is decided to be the class of the test document. The proposed method has been validated on 3 benchmarking datasets viz., 20Newsgroups, Reuters21578 and RCV1-v2 and the results are encouraging.

ACKNOWLEDGEMNT

The second author of this paper acknowledges the financial support rendered by the University of Mysore under UPE grants for the High Performance Computing laboratory.

REFERENCES

[1] Aggarwal C. C., Zhai C X., (2012). Mining text data, Springer, ISBN 978-1-4614-3222-7.
[2] Bharti K. K., Singh P. K., (2015). Hybrid dimension reduction by integrating feature selection with feature extraction method for text clustering. Expert Systems with Applications 42 3105–3114
[3] Cai D., He X. and Han J,. (2005). Document clustering using locality preserving indexing. IEEE TKDE, vol. 17(12), pp. 1624–1637.
[4] Cai, D., and He, X. (2012). Manifold adaptive experimental design for text categorization. IEEE Transactions on Knowledge and Data Engineering 24(4):707–719.
[5] Deerwester S., Dumais S., Landauer T., Furnas G., and Harshman R. (1990). Indexing by latent semantic analysis. Journal of the ASIS, vol. 41(6), pp. 391–407.
[6] Forman G. (2003). An extensive empirical study of feature selection metrics for text classification. JMLR, 3:1289–1306,.
[7] Guru D. S. and Suhil M., (2015). A Novel Term_Class Relevance Measure for Text Categorization. Procedia Computer science, Elsevier, Vol 45, pp. 13-22.
[8] Guyon, I., Elisseeff, A. (2003). An introduction to variable and feature selection, J. Mach. Learn. Res. 3 1157–1182.
[9] Harish B. S., Guru D. S., and Manjunath. S., (2010). Representation and Classification of Text Documents: A Brief Review. IJCA Special Issue on RTIPPR, pp. 110-119.
[10] Hotho A., Nurnberger A. and Paab G., (2005). A brief survey of text mining. Journal for ComputationalLinguistics and Language Technology, vol. 20, pp.19-62.
[11] Javed K., Maruf S., Haroon A B., A two-stage Markov blanket based feature selection algorithm for text classification, Neurocomputing 157 (2015) 91–104.
[12] Li Y. H. and Jain A. K., (1998). Classification of text documents. The Computer Journal, vol. 41(8), pp. 537–546
[13] Montanes E, Diaz I, Ranilla J, Combarro E F., Fernandez J., (2005). Scoring and selecting terms for text categorization. IEEE Intell. Syst. 20 (3) 40–47.
[14] Makrehchi.,(2007). Feature ranking for text classifiers (Ph.D thesis), Dept. of Electrical and Computer Engineering, University of Waterloo, Waterloo, Ontario, Canada.
[15] Rigutini L., (2004). Automatic text processing: Machine learning techniques. Ph.D. Thesis, University of Siena.
[16] Sebastiani F., (2002). Machine learning in automated text categorization. ACM Comput. Surveys, vol. 34 (1), pp. 1–47Uysal A.K., Gunal S.,(2014). Text classification using genetic algorithm oriented latent semantic features, Expert Syst. Appl. 41 (13) 5938–5947.